\documentclass{PoS}
\usepackage{subfigure}
\usepackage{svg}
\usepackage{sidecap}
\usepackage{longtable}
\usepackage{lineno}
\usepackage{ctable}
\usepackage{threeparttable}
\usepackage{verbatim}
\usepackage{tablefootnote}

\title{Ten years of All-sky Neutrino Point-Source Searches.}

\ShortTitle{IceCube Neutrino Sources}

\author{
The IceCube Collaboration\footnote{For collaboration list, see PoS(ICRC2019) 1177.}\\
{\itshape \href{http://icecube.wisc.edu/collaboration/authors/icrc19_icecube}{http://icecube.wisc.edu/collaboration/authors/icrc19\_icecube}}\\
E-mail: \email{tcarver@icecube.wisc.edu}
}

\abstract{

 These proceedings present the results of point-like neutrino source searches using $\sim 10$ yrs of IceCube data from Apr.~6, 2008 to Jul.~10, 2018. 
 We evaluate the significance of an astrophysical signal from a point-like source looking for an excess of clustered neutrino events with energies above $\sim 1$~TeV among the background of atmospheric muons and neutrinos. We perform a full sky scan, a search based on a selected source catalog, and a catalog population study. %, and three stacked galactic catalog searches. 
 The most significant location in the Northern hemisphere from the full-sky scan is compatible with the Seyfert galaxy NGC 1068. This object had also been identified in the source catalog search which finds a $2.9\sigma$ excess after accounting for statistical trials. The combination of this result along with excesses observed at the coordinates of three other sources, including TXS 0506+056, suggests that collectively correlations with sources in the Northern catalog are inconsistent with background at the level of 3.3$\sigma$. These results motivate further interest in such point-like sources which should become observable or ruled out after accumulation of more data or with future detectors. \\

\vspace{4mm}
{\bfseries Corresponding authors:}
\speaker{Tessa Carver}$^{1}$\\
{$^{1}$ \itshape Universit{\'e} de Gen{\`e}ve}
}

\FullConference{36th International Cosmic Ray Conference -ICRC2019-\\
		July 24th - August 1st, 2019\\
		Madison, WI, U.S.A.}

\begin{document}

\section{Introduction}\label{sec:intro}
The origin of high-energy charged particles, known as Cosmic Rays (CRs), reaching up to $\sim10^{20}$~eV has been a mystery for astronomers for over one hundred years~\cite{Hess:1912srp}. The particles themselves are deviated by magnetic fields on their journey to the Earth and as such their origins can only be probed using other messengers such as $\gamma$-rays and neutrinos. Very-high-energy $\gamma$-rays provide evidence for astrophysical acceleration sites but can be produced by both leptonic and hadronic processes and above 1~TeV are attenuated by low-energy photon backgrounds. In comparison, only hadronic processes would produce an astrophysical neutrino flux. Neutrinos would travel unattenuated and undeviated in a direct path from the source. Astrophysical neutrino observations are therefore critical for definitive evidence for the hadronic nature of CR sources or to discover distant very high-energy accelerators. 

IceCube has discovered astrophysical neutrinos in multiple diffuse flux searches \cite{Aartsen:2016xlq,Aartsen:2017mau,Aartsen2013,Aartsen:2014gkd} although point-like sources expected to be responsible for this flux were not resolved~\cite{Abbasi:2010rd,Aartsen:2013uuv,Aartsen:2014cva,Aartsen:2016oji,Aartsen:2018ywr}. %Notably, a potential flaring neutrino source, TXS 0506+056, has since been identified through a multi-messenger campaign around a high-energy IceCube event in Sep.~2017~\cite{eaat1378}. IceCube found additional evidence for flaring neutrino emission from the location of TXS 0506+056 using archival data from 2014-15~\cite{IceCube:2018cha}. Nonetheless, the estimated flux during the 110 day flare cannot account for more than 1\% of the observed diffuse emission \cite{Aartsen:2016xlq}.

Here we present results from several searches for point-like neutrino sources using 10 years of IceCube observations. 
\section{Data Selection}\label{sec:data}
IceCube is a cubic-kilometer neutrino detector installed in the ice at the geographic South Pole~\cite{Aartsen:2016nxy} between depths of 1450\,m and 2450\,m, completed in 2010. Reconstruction of the direction, energy and flavor of the neutrinos relies on the optical detection of Cherenkov radiation emitted by secondary charged particles produced in the interactions of neutrinos in the surrounding ice or the nearby bedrock. This analysis targets astrophysical muon neutrinos ($\nu_\mu$) through their charged-current interactions resulting in a muon which traverses the detector . The majority of the background events in this analysis originate from CRs interacting with the atmosphere to produce cascades of particles including atmospheric muons and neutrinos. The atmospheric muons from the Southern hemisphere are able to penetrate the ice and are detected as track-like events at a rate more than five orders-of-magnitude higher than the corresponding atmospheric neutrinos. However, in the Northern hemisphere almost all of the atmospheric muons are absorbed by the Earth. Despite this physical filtering, poorly-reconstructed atmospheric muons from the Southern sky still constitute a non-negligible background in the Northern hemisphere. The remainder of the background is due to atmospheric $\nu_\mu$, acting as an irreducible background in both hemispheres. 
Neutral current interactions or $\nu_e$ and $\nu_\tau$ charged current interactions produce particle showers which induce spherical cascade-like events. Tracks at $\sim$~TeV energies are reconstructed with a typical median angular resolution of $\lesssim 1^\circ$, while cascades have an angular resolution of $\sim 10^\circ-15^\circ$\cite{Aartsen:2017eiu}. This analysis selects track-like events because of their more precise angular resolution. Tracks have the additional benefit of a greater detectable event rate since they can be observed when the neutrino interaction vertex is located outside of the detector.
\begin{table}
\caption{IceCube configuration, lifetime, number of events, start/ end time and published reference in which the sample selection is described. }
\label{tab:livetimesAbr}
\centering
\begin{tabular}{p{2.5cm}  p{1.2cm}  p{1.6cm} p{1.7cm} p{1.7cm} p{1.8cm}} \hline \\[-5pt]
    \multicolumn{6}{c}{Data Samples} \\[5pt]
    \hline
    Year & lifetime (Days) & Number of Events & Start Day & End Day & Ref.\\[5pt] \hline
    IC40 & 376.4 & 36900 & 2008/04/06 & 2009/05/20 & \cite{Abbasi:2010rd} \\[5pt] \hline
    IC59 & 352.6 & 107011 & 2009/05/20 & 2010/05/31 & \cite{Aartsen:2013uuv} \\[5pt] \hline
    IC79 & 316.0 & 93133 & 2010/06/01 & 2011/05/13 & \cite{Aartsen:2016oji} \\[5pt] \hline
    IC86-2011 & 332.9 & 136244 & 2011/05/13 & 2012/05/15 & \cite{Aartsen:2014cva} \\[5pt] \hline
    IC86-2012-18 & 2198.2 & 760923 & 2012/04/26\footnotemark & 2018/07/10 & See Text \\[5pt]
    %IC86-2012-18 & 2198.2 & 760801 & 2012/05/15 & 2018/07/10 & This work \\[5pt]
    \hline
\end{tabular}
\begin{tablenotes}
\item{[1] start date for test runs of the new processing. The remainder of this run began 2012/05/15}
\end{tablenotes}
\end{table}
%\footnotetext{start date for test runs of the new processing. The remainder of this run began 2012/05/15}

\begin{figure}
    \centering
    \subfigure{\includesvg[width=0.45\textwidth]{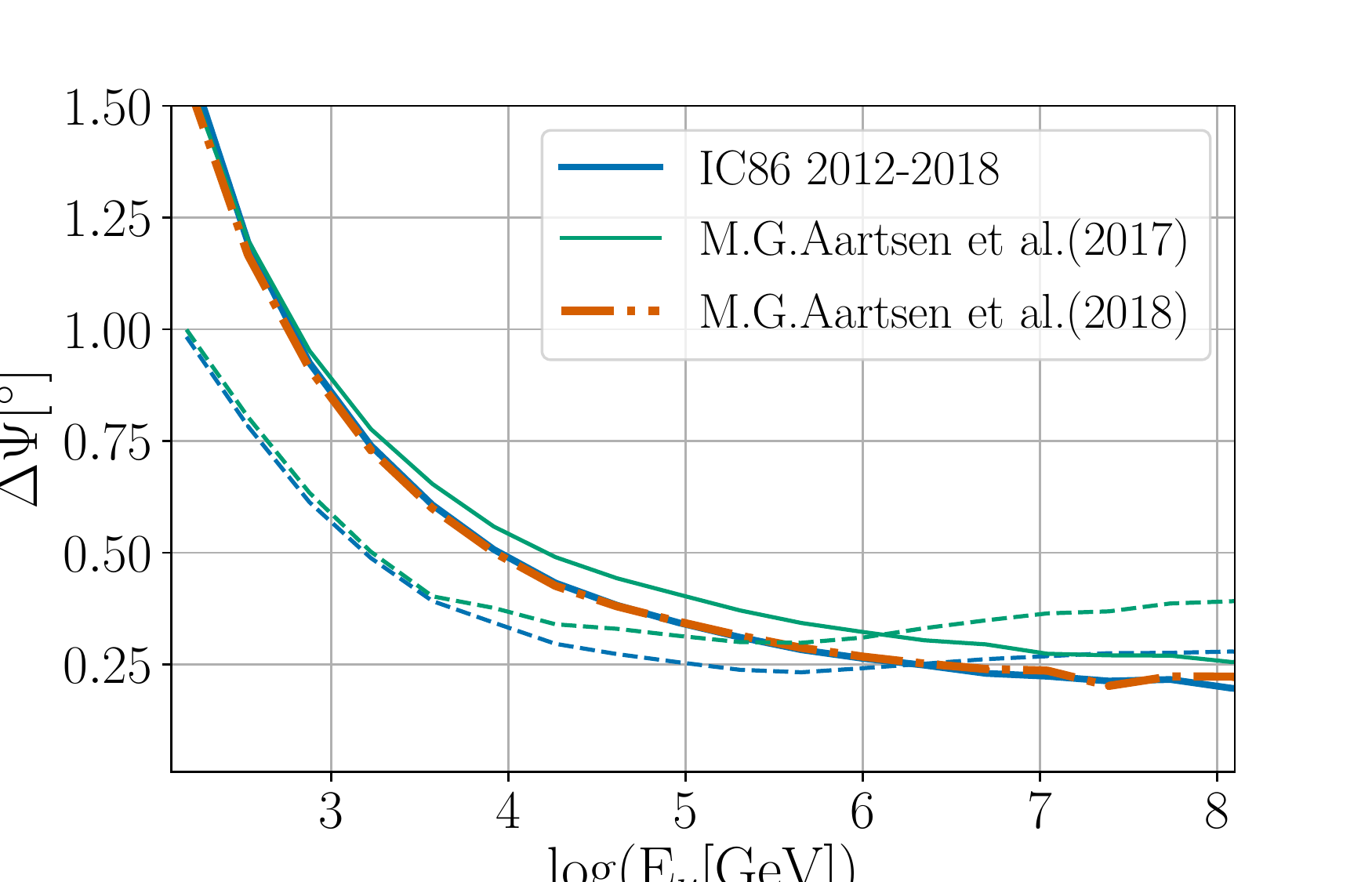}}
    \subfigure{\includesvg[width=0.45\textwidth]{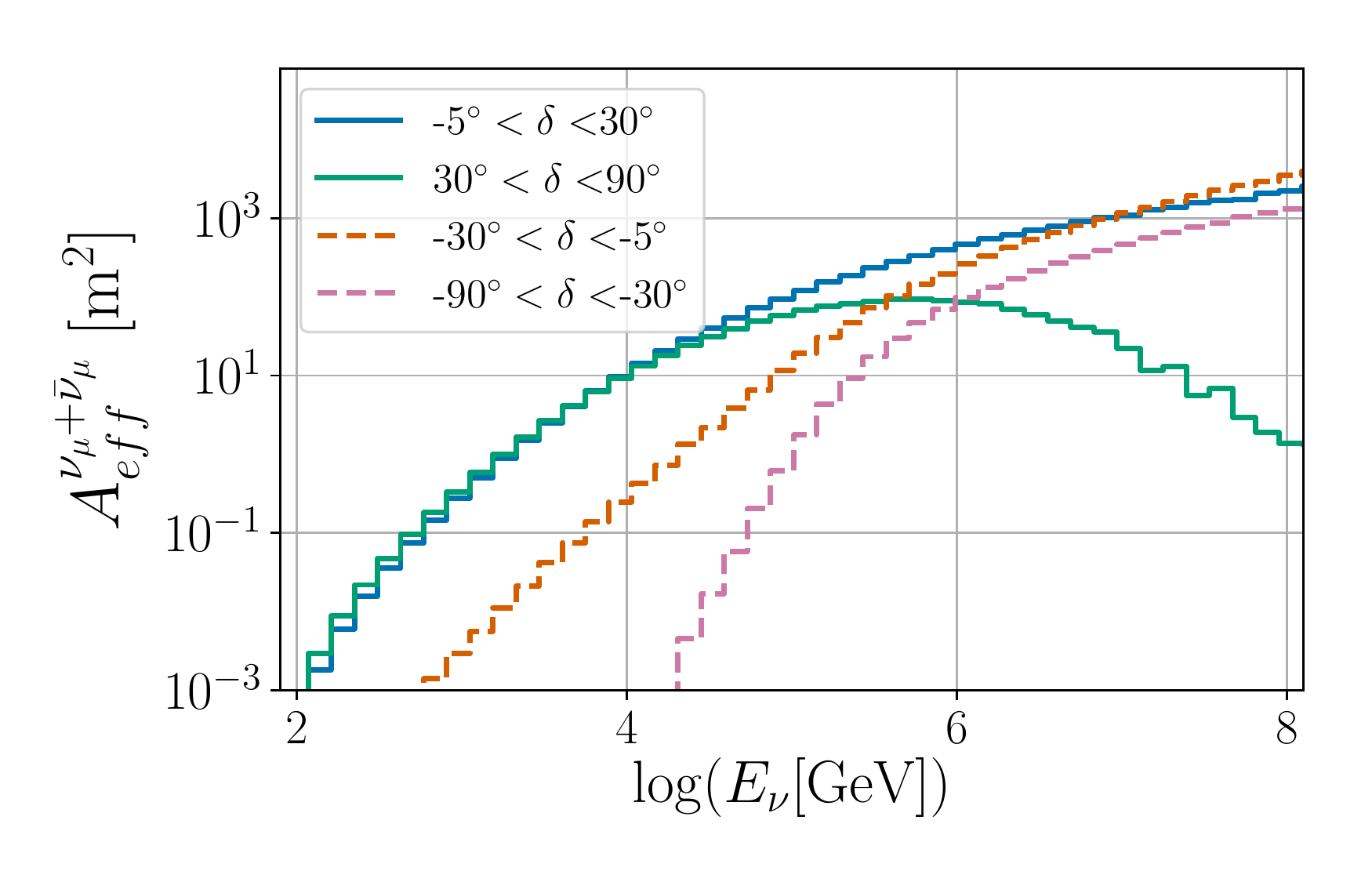}}
    \caption{\textit{Left:} The median angle between simulated neutrinos and corresponding reconstructed muons vs neutrino energy for the data selection used here compared to that in the 7yr all-sky analysis~\cite{Aartsen:2016oji} (solid and dashed lines are for Northern and Southern hemispheres) and in the 8yr Northern sky analysis~\cite{Aartsen:2018ywr}. \textit{Right:} The effective area as a function of neutrino energy for the IC86 2012-2018 event selection in declination bins.}
    \label{fig:psf}
\end{figure}

IceCube began taking data while it was still under construction for three years and functioned with 40, 59, and 79 strings during which the event selection and reconstruction was updated until it stabilized in 2012, as detailed in Table.~\ref{tab:livetimesAbr}. Previously seven years of tracks were analyzed to search for point-sources~\cite{Aartsen:2016oji}. Subsequently, an eight-year sample of tracks from the Northern sky, applied to muon neutrino diffuse flux searches, was also analyzed looking for point-sources~\cite{Aartsen:2018ywr}. The event selection in this work unifies the event filtering adopted in these two past searches. This unification includes updating the direction reconstruction~\cite{Ahrens:2003fg,Aartsen:2013bfa} to use information on the deposited event energy in the detector. This improves the angular resolution by more than 10$\%$ for events above 10~TeV compared to the seven-year study~\cite{Aartsen:2016oji}, and achieves a similar angular resolution to the eight-year Northern diffuse track selection~\cite{Aartsen:2018ywr} (see Fig.~\ref{fig:psf}). 

Track-like events from the Northern and Southern hemisphere are selected using different criteria with a boundary at declination $\delta=-5^\circ$, where the background composition begins to differ significantly. 
Atmospheric muons in the Northern hemisphere can be removed by selecting high-quality track-like topologies. In the Southern hemisphere, the atmospheric background is reduced by strict cuts on the reconstruction quality and minimum energy, since the astrophysical neutrino fluxes are expected to have a harder energy spectrum than the background of atmospheric muons and neutrinos. This effectively removes almost all Southern events with an estimated energy below $\sim10$~TeV.% (see Fig.~\ref{fig:enPDF} in the appendix). 
 
In the Northern hemisphere the multi-variate Boosted Decision Tree (BDT) used to filter events has been updated to a single BDT trained to recognize three classes of events: single muon tracks from atmospheric and astrophysical neutrinos, atmospheric muons, and cascades, where only neutrino-induced tracks are treated as signal. This BDT uses 11 variables related to event topology and reconstruction quality and preserves $\sim90\%$ of the atmospheric neutrinos and $\sim 0.1\%$ of the atmospheric muons from the pre-BDT selection. % of track-like events. 
The Southern hemisphere BDT and selection filters are consistent with the previous seven-year analysis~\cite{Aartsen:2016xlq}. The final event rate of $\sim2~$mHz is dominated by muons from atmospheric neutrino interactions in the Northern hemisphere
and by high-energy well-reconstructed single muons in the Southern hemisphere.

\section{Methods and Results}\label{sec:method}
The point-source searches presented here use a maximum likelihood-ratio technique~\cite{Abbasi:2010rd,Braun:2008bg} to compare the hypothesis of a point-like signal plus diffuse and/or isotropic background versus a background-only null hypothesis. This technique was also applied in the seven and eight-year point source searches~\cite{Aartsen:2016oji,Aartsen:2018ywr}. The all-sky scan and the selected source catalog searches look separately in each hemisphere for a single direction which maximizes this likelihood-ratio. This analysis demonstrates a $\sim35\%$ improvement compared to the seven-year all-sky search~\cite{Aartsen:2016oji} % to a neutrino flux with an $E^{-2}$ spectrum 
due to the longer lifetime and updated event selection. In the Northern hemisphere this analysis is comparable to the eight-year analysis for an $E^{-2}$ spectrum~\cite{Aartsen:2018ywr}, however this work achieves a $\sim30\%$ improvement in sensitivity to sources with a softer spectrum, such as $E^{-3}$. 

\paragraph{All-Sky Scan:}
The expected neutrino emission may not be inferred directly from electromagnetic (EM) observations, as this could vary significantly with the source environment. For this reason, an all-sky search is conducted that is unbiased by EM observations by evaluating the signal-over-background likelihood-ratio at a grid of points across the entire sky. We use a binning of $\sim0.1^\circ \times \sim0.1^\circ$ whilst excluding 8$^\circ$ regions around the celestial poles due to poor statistics. 

At each position on the grid, the likelihood-ratio results in a maximum test-statistic (TS), a best fit number of astrophysical neutrino events ($\hat{n}_s$), and the spectral index for a power-law of the signal energy spectrum ($\hat{\gamma}$). 
A pre-trial p-value is calculated at each grid location by calculating the TS distribution for many background trials and fitting it with a $\chi^2$ distribution.
%A local pre-trial probability (p-value) of obtaining the given or larger TS value at a certain location from only background is estimated at each grid point by fitting the TS distribution from many background trials with a $\chi^2$ function. 
Each background trial is obtained by assigning each event a new right ascension value, uniform across the sky. % therefore removing any clustering signal. 
The hottest spot for this search in each hemisphere is where the local p-value is smallest. The post-trial probability of finding such a hotspot is estimated by comparing the p-value of the hottest spot in the data with a distribution of hottest spots from a large number of background trials.
%\paragraph{Results:} 

\begin{figure}
    \centering
    \subfigure{\includesvg[width=0.45\textwidth]{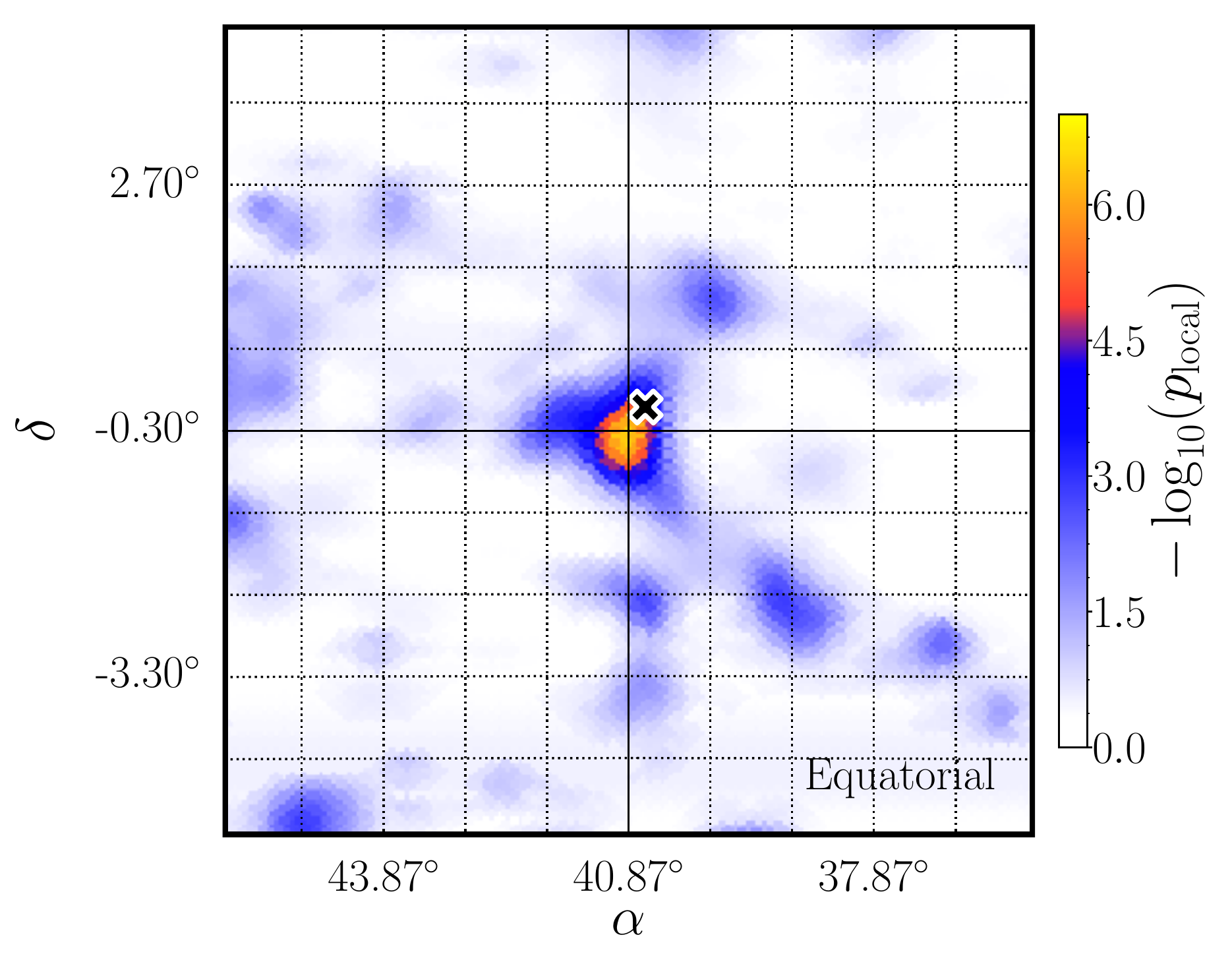}}
    \subfigure{\includesvg[width=0.5\textwidth]{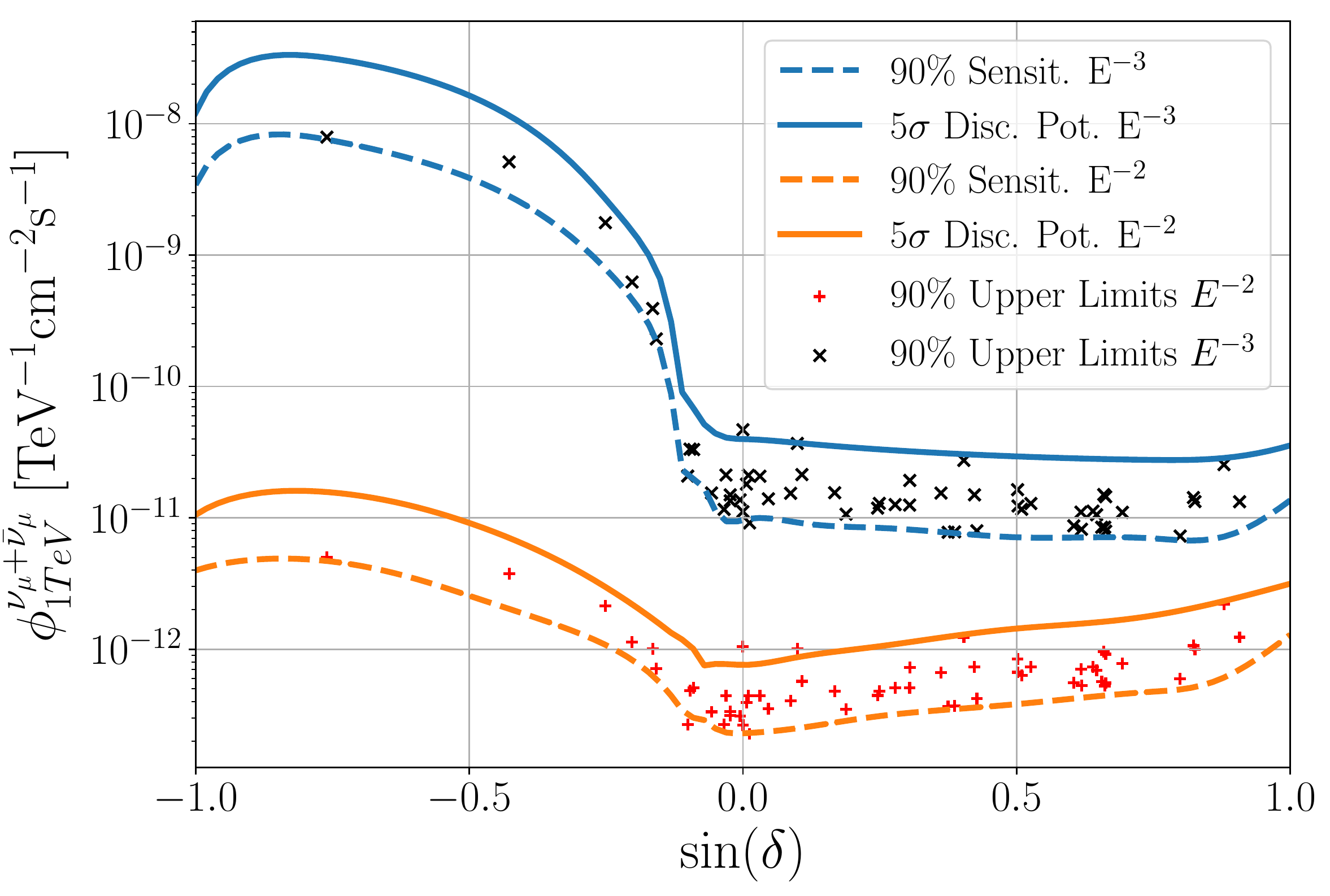}}
    \caption{\textit{Left:} Local pre-trial p-value map around the most significant point in the Northern hemisphere. The black cross marks the coordinates of the galaxy NGC 1068. \textit{Right:} 90$\%$ C.L. mean sensitivity and 5$\sigma$ discovery potential as a function of source declination for a neutrino source with an $E^{-2}$ (orange) and $E^{-3}$ (blue) spectrum. The 90$\%$ upper-limits for the source list are also shown for an  $E^{-2}$ (red) and $E^{-3}$ (black) source spectrum.} 
    \label{fig:nHS}
\end{figure}

The most significant point in the Northern hemisphere is found at equatorial coordinates (J2000) $\alpha=40.9^\circ$, $\delta=$-$0.3^\circ$ with a local p-value of $3.5\times10^{\textrm{-}7}$. The best fit parameters at this spot are $\hat{n}_s=61.5$ and $\hat{\gamma}=3.4$. Due to the large number of points examined, the significance of such a hotspot is 9.9$\%$ when compared to a distribution of largest over-fluctuations in that hemisphere from background trials.
The local probability skymap in a 5$^\circ$ by 5$^\circ$ window around the most significant point in the Northern hemisphere is plotted in Fig.~\ref{fig:nHS}. This point is 0.35$^\circ$ from the coordinates of a galaxy in the Northern source catalog, NGC 1068. 
The most significant point in the Southern hemisphere, at $\alpha=350.2^\circ$, $\delta=\textrm{-}56.5^\circ$, has a pre-trial p-value of $4.3\times10^{\textrm{-}6}$ and fit parameters $\hat{n}_s=17.8$, and $\hat{\gamma}=3.3$. The significance of this hot spot is 75$\%$ post-trial. Both hotspots alone are consistent with a background only hypothesis.

\paragraph{Source Catalog Searches:} 
A catalog composed of 110 sources was constructed which updates the catalog in previous source searches~\cite{Aartsen:2016oji} by using the latest $\gamma$-ray observations. The size of the catalog was chosen to limit the trial factor applied to the most significant source in the catalog such that a source with a 5$\sigma$ local p-value would have at least 4$\sigma$ significance when accounting for the number of sources examined. This list is composed of Galactic and extragalactic sources which are selected separately. 
Extragalactic objects are taken from the third \textit{Fermi}-LAT source catalog (3FGL) catalog~\cite{Acero:2015hja} which provides the highest-energy unbiased measurements of $\gamma$-ray sources over the full sky. Sources 3FGL are weighted according to the product of the integral \textit{Fermi}-LAT flux above 1~GeV and the sensitivity of this analysis at the respective source declination. The 5$\%$ highest-weighted BL Lacs and FSRQs are selected directly and the minimum integral flux from the combined selection of BL Lac and FSRQ sources is used as a flux threshold to include sources marked as unidentified blazars and AGN. Eight \textit{Fermi}-LAT sources are identified as galaxy types with associated starburst activity. Since these types of objects are thought to host hadronic emission~\cite{Loeb:2006tw,Murase:2013rfa}, they are all included in the final source list. 

Galactic sources are included in the catalog by considering measurements of very-high-energy $\gamma$-ray sources from the TeVCat online catalog of $>200$ GeV emitters~\cite{2008ICRCTevCat}. 
Spectra from TeVCat were converted to equivalent neutrino fluxes, assuming a purely hadronic origin behind the observed $\gamma$-ray emission, and compared to the sensitivity of IceCube at the declination of the source (Fig.~\ref{fig:nHS}). Galactic objects with fluxes $>50\%$ of the analysis sensitivity limit were included in the source catalog making a total of 12 Galactic $\gamma$-ray sources.

The final list of neutrino point-source candidates is a Northern-sky catalog containing 97 objects (87 extragalactic and 10 Galactic) and a Southern-sky catalog containing 13 sources (11 extragalactic and 2 Galactic). The large North-South disparity corresponds to the difference in the sensitivity of IceCube in the Northern and Southern hemispheres. The post-trial p-value for each catalog is calculated from the fraction of background trials in which the pre-trial p-value of the most significant source is smaller than the pre-trial p-value found in data.

The 90\% flux upper-limits are shown in Fig.~\ref{fig:nHS} based on the pre-trial p-values for each source, together with the expected sensitivity and discovery potentials.
The most significant excess in the Northern catalog is found in the direction of the galaxy NGC 1068 with a pre-trial p-value of $1.8\times10^{-5}$ (4.1~$\sigma$). The best fit parameters are $\gamma=3.2$ and $\hat{n}_s=50.4$, consistent with the results for the all-sky Northern hottest spot, $0.35^\circ$ away. 
%The parameters of the best fit spectrum at the coordinates of NGC 1068 are shown in Fig.~\ref{fig:nHS}.
When the significance of NGC 1068 is compared to the most significant excesses in the Northern catalog from background trials, the post-trial significance is $2.9\sigma$. To study whether the $0.35^\circ$ offset between the all-sky hotspot and NGC 1068 could be a statistical fluctuation given a soft-spectrum source, we simulate a point source with an $E^{-3.2}$ flux at the \textit{Fermi}-LAT coordinates for NGC 1068 in our background samples. Scanning in a $5^\circ$ window around the injection point, we find the most significant hotspot is identified within $1^\circ$ of the injection point 70\% of the time, and within $0.35^\circ$ of the injection point 51\% of the time. Thus, it is plausible that the offset between the all-sky hotspot and NGC 1068 is a statistical fluctuation.

% \begin{figure}
%     \centering
%     %\includesvg[width=0.5\textwidth]{images/Northern_hotspot.svg}
%     \includegraphics[width=0.5\textwidth]{images/Gamma_Scan_ngc1068_preliminary.png}
%     \caption{Likelihood map at the position of NGC 1068 as a function of the estimated astrophysical flux spectral index and normalisation at 1~TeV. Contours show 1, 2, and 3$\sigma$ confidence intervals assuming Wilks' thoerem with 2 degrees of freedom~\cite{Wilks:1938dza}. Best fit point marked with "x". }
%     \label{fig:gammaScan}
% \end{figure}

The most significant excess in the Southern catalog has a pre-trial p-value of 5.9$\%$ in the direction of PKS 2233-148. The associated post-trial p-value is 55\%, which is consistent with background. 

%From Fig.~\ref{fig:nHS}, the four sources in the Northern catalog with pre-trial p-value of less than 1$\%$ are seen with upper-limits comparable to the 5$\sigma$ discovery potential: NGC 1068, TXS 0506+056, PKS 1424+240, and GB6 J1542+6129. %Evidence has been presented for TXS 0506+056 to be a flaring neutrino source~\cite{IceCube:2018cha} using an overlapping event selection. In this work we find a pre-trial significance of 3.55$~\sigma$ at the coordinates of TXS 0506+056 for a best fit spectra of approximately $E^{-2}$, thus is consistent with those results. 

As in the eight-year point-like source search~\cite{Aartsen:2018ywr}, a source population study is conducted to understand if multiple less significant excesses within the catalog can cumulatively indicate a population of neutrino sources within the catalog. This population study uses the pre-trial p-values already calculated for each source in the catalog and searches for an excess in the rate of small p-values compared to the uniform background expectation. For a catalog of $N$ objects, the cumulative binomial probability ($p_\mathrm{bkg}$) of finding $k$ objects below a given p-value theshold $p_k$ is:

% \begin{equation}
%     p_\mathrm{bkg}=\sum _{i=k}^{N}P_\mathrm{binom}(i|p_k,N)=\sum _{i=k}^{N}\binom{N}{i}p_k^i(1-p_k)^{N-i}\mathrm{ .}
% \end{equation}

The probability threshold ($p_k$) is increased iteratively to vary $k$ between 1 and $N$ sources in order to maximise the sensitivity to any possible population size of true neutrino sources within the catalog. The final result of this search is the most significant $p_{bkg}$ from $N$ different tested values of $k$, hence the post-trial p-value from this search takes into account a trial factor for the different tested values of $k$. 

The most significant pre-trial p-value from the Northern catalog population analysis is $3.3 \times 10^{\textrm{-}5}$ (4$\sigma$) which is found for $k=4$. The resulting flux upper limits of the four most significant Northern hemisphere sources contributing to this excess with p-value less than 1\% (NGC 1068, TXS 0506+056, PKS 1424+240, and GB6 J1542+6129) are comparable to the 5$\sigma$ discovery potentials (see Fig.~\ref{fig:nHS}). When accounting for the fact that different signal population sizes are tested, the post-trial p-value is $4.8 \times 10^{\textrm{-}4}$ (3.3$\sigma$).
%four most significant sources discussed above: NGC 1068, TXS 0506+056, and the BL Lacs PKS 1424+240 and GB6 J1542+6129. 
Since evidence has already been presented for TXS 0506+056 to be a flaring neutrino source~\cite{IceCube:2018cha}, an a-posteriori search is conducted removing this source from the catalog resulting in a 2.25$\sigma$ post-trial excess due to the remaining 3 most significant sources. 
For the Southern catalog the most significant excess is 12\%, provided by 5 of the 13 sources. The resulting post-trial p-value is 36\%, consistent with background.
\begin{comment}
\paragraph{Stacked Source Searches}
Where certain classes of sources are expected to produce similar fluxes, 
stacking searches require a lower flux per source for a discovery than considering each source individually. 
Three catalogs of $\gamma$-ray Galactic sources were constructed to be stacked in this work.
Sources are selected from TeVcat and categorized into pulsar wind nebula (PWN), supernova remnants (SNR) and unidentified objects (UNID), with the aim of grouping objects likely to have similar properties as neutrino emitters. The final groups consist of 33 PWN, 23 SNR, and 58 UNID. The relative contribution expected from each source within a catalog is estimated using the integral $\gamma$-ray flux above 10~TeV. 
\end{comment}
%\paragraph{Results:}
Table~\ref{tab:results_summary} includes the summary of all search results.
%, including the p-values from these 3 stacking searches, which are consistent with background fluctuations. 
%All three catalogs are considered consistent with the background hypothesis. 

\begin{table}
\caption{Summary of final p-values (pre-trial and post-trial) for each point-like source search implemented in this paper.}
\label{tab:results_summary}
\centering
%\label{tab:stacking_results}
\begin{tabular}{ p{2.5cm}  p{1.5cm}  p{1.9cm}  p{3.5cm}  }\hline \\[-5pt]
    %\multicolumn{4}{c}{Results Summary} \\
    %\hline
    Analysis    &  Category  & Best $p_{local}$ &  Post-trial significance \\
    \hline \hline
    All-Sky Scan &  North & $3.5\times10^{-7}$ & $9.9\times10^{-2}$ \\
    &  South & $4.3\times10^{-6}$ & 0.75 \\
    \hline
    Source List &  North & $1.8\times10^{-5}$ & $2.0\times10^{-3}$ (2.9$\sigma$)\\
    &  South & $5.9\times10^{-2}$ & 0.55 \\
    \hline
    Catalog  & North & 3.3$\times10^{-5}$ & $4.8\times10^{-4}$ (3.3$\sigma$) \\
    Population & South & 0.12 & 0.36 \\
    %\hline
    %Stacking  & SNR & -- & 0.11 \\
    %Search& PWN & -- & 1.0 \\
    %& UNID & -- & 0.4 \\
    \hline
\end{tabular}
\end{table}

\section{Conclusions}\label{sec:conclusion}
An updated event selection optimized for neutrino point-like sources is applied to 10 years of IceCube data taken from April 2008 to July 2018. Multiple methods to search for neutrino point-sources were used: an unbiased all-sky scan, and a source catalog searches and corresponding catalogue population studies for each hemisphere. The results, summarised in Table~\ref{tab:results_summary}, show no evidence for an astrophysical neutrino source in the Northern and Southern hemispheres. However, the most significant source in the Northern catalog, NGC 1068, shows a 2.9$\sigma$ excess at the \textit{Fermi} coordinates when accounting for statistical trials. These coordinates are 0.35$^\circ$ from the most significant point in the Northern hemisphere. The entire Northern source catalog provides a 3.3$\sigma$ inconsistency with a background only hypothesis in a population analysis from an excess of significant p-values in the directions of the Seyfert II galaxy NGC 1068, the blazars TXS 0506+056, PKS 1424+240, and GB6 J1542+6129.

NGC 1068 is the most luminous Seyfert II galaxy observed by \textit{Fermi}-LAT~\cite{Ackermann:2012vca} at a 14.4 Mpc distance, and is a known site of star-formation. 
While the estimated neutrino flux from this work appears higher than current neutrino models inferred from \textit{Fermi}-LAT measurements, the large uncertainty from our spectral measurement and the high X-ray and $\gamma$-ray absorption along the line of site~\cite{2015A&A...584A..20W,Lamastra:2017iyo} prevent a straight forward connection.
After 10 years of astrophysical neutrino source searches, the results presented here demonstrate that some of the first point-like sources could potentially be emerging from the background. More years of data would improve the sensitivity of this analysis and reconstruction improvements and increased statistics from IceCube Gen-2~\cite{vanSanten:2017chb} are estimated to improve the discovery potential of the point source analysis by an order of magnitude. This would make the analysis sensitive to a flux lower than the best fit neutrino flux of the four most significant sources from this analysis, including NGC 1068. 
%Thus, these four sources are likely to be either confirmed as astrophysical neutrino sources or ruled out at a 90\% C.L. given more years of data or a detector upgrade.  

%\section{Listing some References}\label{sec:refs}
% This is a paper from a previous ICRC \cite{Zoll:2015wcu}. This is a second paper from a previous ICRC \cite{Peiffer:2017vsm}. This is a paper from the current ICRC \cite{Hussain:2019icrc_gw}.
% Here is an IceCube journal paper \cite{Aartsen:2016nxy} and an external journal paper \cite{Waxman:1998yy}.

% Set up the bibliography using BibTeX.
% Get references from inspirehep.net or NASA/ADS and put them in references.bib.
\bibliographystyle{ICRC}
\bibliography{references}

% Or, set up the bibliography manually, if you prefer to do things this way.
%
% \begin{thebibliography}{99}
%   \bibitem{Zoll:2015wcu}{{\bf IceCube} Collaboration, \pos{PoS(ICRC2015)1099} (2016).}
%   \bibitem{Peiffer:2017vsm}{{\bf IceCube-Gen2} Collaboration, \pos{PoS(ICRC2017)1052} (2018).}
%   \bibitem{Hussain:2019icrc_gw}{{\bf IceCube} Collaboration, \pos{PoS(ICRC2019)xyz} (these proceedings).}
%   \bibitem{Aartsen:2016nxy}{{\bf IceCube} Collaboration, M.~G.~Aartsen {et al.}, \emph{JINST} {\bf 12} (2017) P03012%
%   % optionally add arXiv ID here [{\tt astro-ph/1612.05093}]
%   .}
%   \bibitem{Waxman:1998yy}{E. Waxman and J. N. Bahcall, \emph{Phys. Rev.} {\bf D59} (1999) 023002.}
% \end{thebibliography}
%\input{appendix.tex}
\end{document}